\documentclass[
superscriptaddress,nofootinbib]{revtex4}
\usepackage{epsfig}
\usepackage[english]{babel}
\usepackage{pgf}
\usepackage{amsfonts,amsmath,amssymb,enumerate,array,amssymb}
\usepackage{enumitem}
\usepackage{mathtools} 
\usepackage{multirow} 
\usepackage[latin1]{inputenc}
\usepackage{bm}
\usepackage{fixmath}
\usepackage{amsbsy}
\usepackage{graphicx}
\usepackage{capt-of}
\usepackage{dcolumn}
\usepackage{upgreek}
\usepackage{float}
\usepackage{subfig}
\usepackage[format=plain,justification=centerlast]{caption}
\usepackage{bbm}        

\usepackage{layouts}
\usepackage{natbib}
\usepackage{abrev-jour-long}

\usepackage{pdfpages}
\usepackage{color}
\usepackage{graphicx,epstopdf}
\usepackage[colorlinks=true,
linkcolor=red,
urlcolor=purple,
citecolor=blue,hyperindex]{hyperref}

\begin{document}

	\title{Asymmetrical braneworlds and the charged lepton mass spectrum}

	\author{Henrique Matheus Gauy}\email[]{henmgauy@df.ufscar.br}

	\author{Alex E. Bernardini}\email[]{alexeb@ufscar.br}

	\affiliation{Departamento de F\'isica, Universidade
		Federal de S\~{a}o Carlos, S\~ao Carlos, 13565-905 SP, Brazil}

	
	\begin{abstract}
		A braneworld mechanism for explaining the mass spectrum of the charged leptons is proposed. Based on the existence of an asymmetric warp factor for a $5+1$-dim braneworld scenario, the proper fractions between the masses of the electron, muon and tauon are achieved. As a straightforward consequence, our results coincide with the Koide's mass formula.
	\end{abstract}
	

	\date{\today}
	
	\maketitle
	
\textit{Introduction} - Within the Standard Model (SM), the masses and mixings of the quarks and leptons originates from their interactions with the Higgs field. Even though such interactions have been experimentally confirmed, the interaction coupling constants are free parameters and the generating mechanism of the relations between them is still unveiled. Notwithstanding the well-defined mass spectrum exhibited by the three families of charged leptons, an explanation for the mass values and their relative gaps is indeed an open problem. 
Phenomenological approaches \cite{Koide,Sumino} have been proposed through empirical relations among the fermion masses, as an attempt of uncovering some of its underlying physics\footnote{For instance, the so-called Koide's mass formula \cite{Koide,Sumino},
\begin{equation}
\mathcal{K}=\frac{m_{e}+m_{\mu}+m_{\tau}}{\left(\sqrt{m_{e}}+\sqrt{m_{\mu}}+\sqrt{m_{\tau}}\right)^{2}}=\frac{2}{3},\label{koide}
\end{equation}
provides such a speculative relation which can be translated as a weaker condition for the fractions $m_{\mu}/m_{e}\approx 207$ and $m_{\tau}/m_{\mu}\approx 17$, here $m_{e}$, $m_{\mu}$ and $m_{\tau}$.}.

In a parallel context, extra dimensions have played a prominent role in our understanding of the hierarchy between the Planck and weak scales \cite{RS-1,ArkaniHamed1998a}. Thus, it is natural to assume that other properties of the SM could also be understood from such paradigm. The mass spectrum of fermions should be no different. The most promising higher dimensional scenarios are based on braneworld models with non-factorizable geometries \cite{RS-1,RS-2,Gauy2022,Gauy2022-b,Bernardini2016,Almeida2009,Bazeia2004,Dzhunushaliev2010,Bazeia2002,DeWolfe2000,Ahmed2013,Gremm2000a,Kehagias2001,Kobayashi2002,Bronnikov2003,BAZEIA2009b,BarbosaCendejas2013,ZHANG2008,Melfo2003,Bazeia2009a,Koley2005,Bazeia2014,Chinaglia2016,Bernardini2013,Rosa}, where a $\mathbb{Z}_{2}$-symmetric brane is mostly assumed\footnote{Since they are generally motivated by the Horava-Witten model \cite{HORAVA1,HORAVA2}.}. 
Nevertheless, generalization of these models are obtained by relaxing the mirror symmetry across the brane \cite{Bazeia2014,Kraus1999,Ida2000,Deruelle2000,Perkins2001,Carter2001,Gergely2003,PhysRevD.76.124009,SHTANOV2008,Charmousis2007,Bowcock2000,Koyama2009,Guerrero2005,PhysRevD.71.024032,Padilla2005,Padilla2005a}. The term ``asymmetric'' brane refers to any braneworld model for which the mirror symmetry is not required. Here, an asymmetric brane model will be an essential feature for realizing the spectrum of the fermions.

In this letter, a braneworld mechanism for explaining the charged lepton mass spectrum is evaluated. Modeling the fermion spectrum through extra dimensions is indeed not a new idea. It has been addressed in the literature through different contexts \cite{ArkaniHamed2000a,ArkaniHamed2001,Dvali2000,Kaplan2000,Huber2001,Frere2001,Frere2001a,Frere2003,Frere2013,Libanov2001,Libanov2002}. However, instead of either considering that the distinct chiralities are differently placed over the extra dimensions \cite{ArkaniHamed2000a,ArkaniHamed2001} or relying on a non-trivial higher dimensional Higgs and several other fields \cite{Dvali2000,Kaplan2000,Huber2001,Frere2001,Frere2001a,Frere2003,Frere2013,Libanov2001,Libanov2002}, a simpler mechanism shall be here admitted. By considering a six-dimensional braneworld constructed from an asymmetric conformally flat metric, a non-trivial bulk profile for the gauge boson and a dark scalar field, fermionic fields, whose dynamics are driven by an ordinary $SU(2)\times U(1)$ action, are shown to give rise to several massive four-dimensional spinors. In particular, for the right choice of the parameters, the number of massive spinors becomes exactly three, and their mass spectrum shall coincide with that for the charged leptons.

The setup for the proposed mechanism comes from a six-dimensional braneworld $\mathbb{E}^{6}$ that is, as a set, equivalent to the product
space $\mathbb{M}^{4}\times \mathbb{R}\times \mathbb{S}^{1}$, where $\mathbb{M}^{4}$ is some four dimensional pseudo-Riemannian manifold, $\mathbb{R}$ is the real line and $\mathbb{S}^{1}$ is the circle. The following ansatz is assumed for the metric of $\mathbb{E}^{6}$,
\begin{equation}
\boldsymbol{g}=e^{-2A\left(y\right)}\left(\eta_{\mu\nu}\mathrm{d}x^{\mu}\mathrm{d}x^{\nu}+r^{2}\mathrm{d}\theta^{2}+\rho^{2}\mathrm{d}y^{2}\right),\label{bundlemetric}
\end{equation}
where $A$ is the warp factor, $\eta_{\mu\nu}$ is the Minkowski metric\footnote{Clarifying the notation, Greek indices ($\mu$, $\nu$,...) are valued in the set $\{0,1,2,3\}$, uppercase Latin indices ($M$, $N$,...) are valued in $\{0,1,2,3,4,5\}$, lowercase Latin indices ($m$, $n$, $i$, $j$,...) are valued in $\{4,5\}$ (and represent the bulk co-dimensions) and the labels $x^{4}=\theta$ and $x^{5}=y$, represent the choice of coordinates for the co-dimensions $(\mathbb{B}^{2})$; finally, tensors when being referred to its (abstract) entirety will be in boldface, as $\boldsymbol{g}$, but its components will be cast in regular font, as $g_{\mu\nu}$.} of the space-time $\mathbb{M}^{4}$, $\theta\in\mathbb{S}^{1}$, $r$ is the radius of $\mathbb{S}^{1}$, $y\in\mathbb{R}$ and $\rho$ is the brane model scale. An asymmetry of the braneworld is achieved by imposing that $e^{-A}=f_{+}+f_{-}$, where $f_{+}$ and $f_{-}$ are even and odd non-null functions, respectively, and $f_{+}\geq \left|f_{-}\right|\text{ for all }y$.

The fermions are thus represented by the action \cite{Schwartz}
\begin{align}
 S_{D}=&\int\mathrm{d}^{6}x\sqrt{-g}\,\Bigg\{\overline{\mathsf{L}}^{\left(6\right)}{\Gamma}^{M}\Bigg[\nabla_{M}-\left(i\mathsf{g}\tau_{a}\mathsf{W}^{a}_{M}-\frac{i}{2}\mathsf{g}'\mathsf{B}_{M}\right)\Bigg]\mathsf{L}^{\left(6\right)}+\overline{\mathsf{e}}^{\left(6\right)}_{-}{\Gamma}^{M}\Bigg[\nabla_{M}+\left(i\mathsf{g}'\mathsf{B}_{M}+\zeta_{,M}\right)\Bigg]\mathsf{e}^{\left(6\right)}_{-}\nonumber
\\
&\qquad\qquad\qquad\qquad\qquad\qquad\qquad\qquad\qquad\qquad\qquad\qquad\qquad\qquad\qquad-m_{0}\left[\overline{L}^{\left(6\right)}\mathsf{H}\,\mathsf{e}^{\left(6\right)}_{-}+\overline{\mathsf{e}}^{\left(6\right)}_{-}\mathsf{H}^{\dagger}\,L^{\left(6\right)}\right]\Bigg\},\label{chirallepton}
\end{align}
where $\nabla_{M}:=\partial_{M}+{\mathfrak{C}}_{M}$, ${\mathfrak{C}}_{M}$ is the spin connection of $\left(\mathbb{E}^{6},\boldsymbol{g}\right)$, $\zeta$ is some dark scalar field, $\mathsf{B}_{M}$ is the hypercharge gauge boson, $\mathsf{W}^{a}_{M}$ are the $SU(2)$ gauge bosons, $\mathsf{g}$ and $\mathsf{g}'$ are the $SU(2)$ and $U(1)$ couplings, $\tau_{a}=\sigma_{a}/2$ are the $SU(2)$ generators, $\mathsf{H}=\begin{Bmatrix}
	\mathsf{H}_{+}\\\mathsf{H}_{0}
	\end{Bmatrix}$ is the Higgs doublet, and $m_{0}$ is the coupling constant with the Higgs field. The left-handed leptons\footnote{One chooses a representation such that $\Gamma^{7}=\begin{bmatrix}
		I^{4}& 0^{4}\\ 0^{4}&-I^{4}
		\end{bmatrix}$.} pair up to transform under $SU(2)$,
\begin{equation}
\mathsf{L}^{(6)}=\begin{Bmatrix}
\nu^{(6)}_{+}\\\mathsf{e}^{(6)}_{+}
\end{Bmatrix},
\end{equation}
where \begin{equation}
		\nu^{(6)}_{+}=\begin{bmatrix}
		\Psi^{L+}_{\nu}\\ \Psi^{R+}_{\nu}\\0\\0
	\end{bmatrix}\text{ and }\mathsf{e}^{(6)}_{+}=\begin{bmatrix}
	\Psi^{L+}_{\mathsf{e}}\\ \Psi^{R+}_{\mathsf{e}}\\0\\0
\end{bmatrix}
\end{equation} 
represent the left-handed charged leptons and neutrinos, respectively. While right-handed leptons\footnote{Right-handed neutrinos have been disregarded from the formalism since the interest in this work is in the charged lepton masses.}, which are uncharged under $SU(2)$, are represented by
\begin{equation}
\mathsf{e}^{(6)}_{-}=\begin{bmatrix}
0\\0\\\Psi^{L-}_{\mathsf{e}}\\ \Psi^{R-}_{\mathsf{e}}
\end{bmatrix}.
\end{equation}

The lepton masses are a consequence of the existence of a Higgs field, which is driven by the action
\begin{equation}
S_{H}=\int\mathrm{d}x^{6}\sqrt{-g} \left\{\left[\left(\nabla_{M}-i\mathsf{g}\tau_{a}\mathsf{W}^{a}_{M}-\frac{i}{2}\mathsf{g}'\mathsf{B}_{M}\right) \mathsf{H}\right]^{\dagger}\left(\nabla^{M}-i\mathsf{g}\tau^{b}\mathsf{W}_{b}^{M}-\frac{i}{2}\mathsf{g}'\mathsf{B}^{M}\right)\mathsf{H}-\zeta^{,M}\zeta_{,M}\mathsf{H}^{\dagger}\mathsf{H}+V(\mathsf{H})\right\},\label{higgs}
\end{equation}
where $V(\mathsf{H})=\mu^{2}\mathsf{H}^{\dagger}\mathsf{H}-\lambda\left(\mathsf{H}^{\dagger}\mathsf{H}\right)^{2}$.

For conformally flat metrics (cf. \eqref{bundlemetric}), fermionic fields can not be localized at the vicinity of the brane. Hence, our proposal thus relies on the existence of a non-trivial bulk profile for the hypercharge gauge boson $\boldsymbol{\mathsf{B}}=\mathsf{B}_{M}\mathsf{d}x^{M}$ and the scalar field\footnote{The scalar field is not essential for the mechanism, however, simpler expressions and more elegant properties are obtained for the reduced resulting effective four dimensional action in that case.} $\zeta$, each defined by
\begin{equation}
\mathsf{B}_{\mu}=\mathsf{B}_{\mu}\left(x^{\nu}\right)\text{, } \mathsf{W}^{a}_{\mu}=\mathsf{W}^{a}_{\mu}\left(x^{\nu}\right)\text{, }\mathsf{B}_{y}=\mathsf{W}^{a}_{i}=0\text{, }\mathsf{B}_{\theta}=\frac{r}{\rho\mathsf{g}'} \frac{F_{,y}}{F}\text{ and }\zeta=\frac{1}{2}\ln\left(F\right),\label{gaugefield}
\end{equation}
where $F$ is some positive even function of $y$ and the subscript index ``,'' stands for partial derivatives. Gauge and scalar fields defined in Eq. \eqref{gaugefield} drive the localization of ferminonic modes\footnote{With the exception of $\zeta^{,M}\zeta_{,M}\mathsf{H}^{\dagger}\mathsf{H}$, which serves the purpose of achieving a trivial bulk profile for the Higgs, as shall be presented later.} \cite{Liu2014,Liu2007}, and can be interpreted as background fields. 
Their corresponding chiralities are distinguished at the action, in the exact same way as they are identified for four dimensional models. The six dimensional models, as constructed here, do not provide any required structure for distinguishing chiralities by the localization mechanism, which is indeed unnecessary in this case.

The parameters $\mu^{2}$, $\lambda$ and $m_{0}$ set the scale of the charged lepton masses, which are much smaller than the brane model scale $1/\rho$, and are treated as a perturbation of the system. Therefore the terms $\mu^{2}$, $\lambda$ and $m_{0}$ do not affect the co-dimensional wave functions, which are determined as if the fermions and the Higgs were massless. The zero modes of leptons are thus described by the equations
\begin{equation}
\Gamma^{i}\left(\nabla_{i}+i\mathsf{g}'\frac{1}{2}\mathsf{B}_{i}\right)\mathsf{L}^{(6)}_{0}=0,\label{leftleptonzero}
\end{equation}
and
\begin{equation}
\Gamma^{i}\left(\nabla_{i}+i\mathsf{g}'\mathsf{B}_{i}+\zeta_{,i}\right)\mathsf{e}^{\left(6\right)}_{-0}=0.\label{rightchargedleptonzero}
\end{equation}
Following a separation of variables technique, Eqs.~\eqref{leftleptonzero}~and~\eqref{rightchargedleptonzero} are reduced to Schr\"odinger-like equations and the fermionic zero modes are described by\footnote{The components  $\nu^{(6)}_{+0}=e^{\frac{5A}{2}}\sum_{k}R^{+}_{0k}\,\begin{bmatrix}
	0\\ \psi^{R+}_{\nu 0k}(x^{\mu})\\0\\0
	\end{bmatrix}\text{, }\mathsf{e}^{(6)}_{+}=e^{\frac{5A}{2}}\sum_{k}R^{+}_{0k}\,\begin{bmatrix}
	0\\ \psi^{R+}_{\mathsf{e} 0k}(x^{\mu})\\0\\0
	\end{bmatrix}\text{ and }\mathsf{e}^{(6)}_{-0}=e^{\frac{5A}{2}}\sum_{k}L^{-}_{0k}\begin{bmatrix}
	0\\0\\\psi^{L-}_{\mathsf{e}0k}(x^{\mu})\\0
	\end{bmatrix}$ are non-normalizable, since $L^{-}_{0k}={A}_{k}e^{-ik\theta}e^{\frac{k\rho y}{r}}F^{-\frac{3}{2}}$ and $R^{+}_{0k}={B}_{k}e^{-ik\theta}e^{\frac{k\rho y}{r}}F^{-\frac{1}{2}}$.}
\begin{equation}
\nu^{(6)}_{+0}=e^{\frac{5A}{2}}\sum_{k}L^{+}_{0k}\,\begin{bmatrix}
\psi^{L+}_{\nu 0k}(x^{\mu})\\ 0\\0\\0
\end{bmatrix}\text{, }\mathsf{e}^{(6)}_{+}=e^{\frac{5A}{2}}\sum_{k}L^{+}_{0k}\,\begin{bmatrix}
\psi^{L+}_{\mathsf{e} 0k}(x^{\mu})\\ 0\\0\\0
\end{bmatrix}\text{ and }\mathsf{e}^{(6)}_{-0}=e^{\frac{5A}{2}}\sum_{k}R^{-}_{0k}\begin{bmatrix}
0\\0\\0\\\psi^{R-}_{\mathsf{e}0k}(x^{\mu})
\end{bmatrix},\label{fermionzeromode}
\end{equation}
where
\begin{equation}
L^{+}_{0k}={C}_{k}e^{ik\theta}e^{\frac{k\rho y}{r}}F^{\frac{1}{2}}\text{ and }R^{-}_{0k}={D}_{k}e^{ik\theta}e^{\frac{k\rho y}{r}}F^{\frac{1}{2}},\label{wavefunctionlepton}
\end{equation}
represent the fermionic co-dimensional wave functions, $k\in\mathbb{Z}$, and ${C}_{k}$ and $D_{K}$ are integration constants.

On the other hand, the zero mode of the Higgs field satisfies the equation
\begin{equation}
\nabla^{i}\nabla_{i}\mathsf{H}-i\frac{\mathsf{g}'}{r^{2}}\mathsf{B}_{\theta}\mathsf{H}_{,\theta}=0,
\end{equation}
which after a rescaling and a separation of variables technique, with $\mathsf{H}=\sum_{k}e^{ik\theta}e^{2A}\hat{\phi}_{k}H_{k}\left(x^{\mu}\right)$, reduces to
\begin{equation}
\hat{\phi}_{k,vv}+2\left(A_{,vv}-2A_{,v}{}^{2}\right)\hat{\phi}_{k}+k\frac{\mathsf{g}'}{r^{2}}\mathsf{B}_{\theta}\hat{\phi}_{k}=\frac{k^{2}}{r^{2}}\hat{\phi}_{k}.\label{generalscalar}
\end{equation}
The zero mode of scalar fields are generally like $\hat{\phi}_{0}=c \,e^{-2A}$, which is the solution of Eq. \eqref{generalscalar} for\footnote{Other values of $k$ necessarily imply in a massive mode.} $k=0$, implying in a trivial bulk profile for the Higgs field,
\begin{equation}
\mathsf{H}=c\,H_{0}\left(x^{\mu}\right)=c\,\begin{Bmatrix}
\phi_{+}\left(x^{\mu}\right)\\\phi_{0}\left(x^{\mu}\right)
\end{Bmatrix},\label{higgsprofile}
\end{equation}
where $c$ is an integration constant. The existence of the perturbation $m_{0}$ breaks the degeneracy of the zero modes, implying in a tower of spinors with varying masses driven by integer values, $k$. The SM leptons are thus represented by the localizable zero modes and a four-dimensional observer will model them by an effective action\footnote{The mass matrix of the charged leptons is diagonal because the wave functions of the zero modes are orthogonal for different values of $k$.}, which follows from substituting Eqs.~\eqref{fermionzeromode} and~\eqref{higgsprofile} into Eq.~\eqref{chirallepton},
\begin{align}
S_{D}^{\left(4\right)}=&\sum_{k}\int\mathrm{d}^{4}x\Bigg\{i\,\overline{\mathsf{L}}_{k}\,\gamma^{\mu}\Bigg[\nabla_{\mu}-\left(i\mathsf{g}\tau_{a}\mathsf{W}^{a}_{\mu}-\frac{i}{2}\mathsf{g}'\mathsf{B}_{\mu}\right)\Bigg]\,\mathsf{L}_{k}+i\,\overline{\mathsf{e}}_{Rk}\gamma^{\mu}\Bigg(\nabla_{\mu}+i\mathsf{g}'\mathsf{B}_{\mu}\Bigg)\mathsf{e}_{Rk}\nonumber
\\
&\qquad\qquad\qquad\qquad\qquad\qquad\qquad\qquad\qquad\qquad\qquad\qquad\qquad\qquad\quad-m_{k}\left[\,\overline{\mathsf{L}}_{k}\,H_{0}\,\mathsf{e}_{Rk}+\,\overline{\mathsf{e}}_{Rk}\,H^{\dagger}_{0}\,\mathsf{L}_{k}\right]\Bigg\},
\end{align}
where
\begin{equation}
 \mathsf{L}_{k}=\begin{Bmatrix}
 \nu_{Lk}\\\mathsf{e}_{Lk}
 \end{Bmatrix}\text{, }
 \nu_{Lk}=\begin{bmatrix}
 \psi^{L+}_{\nu 0k}\\0
 \end{bmatrix}\text{, }\mathsf{e}_{Lk}=\begin{bmatrix}
 \psi^{L+}_{\mathsf{e} 0k}\\0
 \end{bmatrix}\text{, }\mathsf{e}_{Rk}=\begin{bmatrix}
 0\\ \psi^{R-}_{\mathsf{e}0k}
 \end{bmatrix}
\end{equation}
and
\begin{equation}
 \mathsf{m}_{k}=\frac{
 	\displaystyle m_{0}\int\mathrm{d}y\,e^{-A}\overline{{L}^{+}_{0k}}{R}^{+}_{0k}}{\sqrt{\displaystyle\int\mathrm{d}y\,\left|{L}^{+}_{0k}\right|^{2}}\sqrt{\displaystyle\int\mathrm{d}y\,\left|{R}^{+}_{0k}\right|^{2}}\sqrt{\displaystyle\int\mathrm{d}y\,{\hat{\phi}_{0}}^{2}}}=\frac{m_{0}\displaystyle\int\limits^{\infty}_{-\infty}\mathrm{d}y\,\left(f_{+}+f_{-}\right)Fe^{\frac{2\rho k}{r}y}}{\displaystyle\int\limits^{\infty}_{-\infty}\mathrm{d}y\,Fe^{\frac{2\rho k}{r}y}\sqrt{\displaystyle\int\limits_{-\infty}^{\infty}\mathrm{d}y\,\left(f_{+}+f_{-}\right)^{4}}}.\label{massfermions}
 \end{equation}
 
Analogously, a four dimensional observer will model the Higgs field by an effective action, which follows from substituting Eq.~\eqref{higgsprofile} into Eq.~\eqref{higgs},
\begin{equation}
S^{(4)}_{H}=\int\mathrm{d}x^{4}\left\{\left[\left(\nabla^{\mu}-i\mathsf{g}\tau^{a}W_{a}^{\mu}-\frac{i}{2}\mathsf{g}'B^{\mu}\right) H_{0}\right]^{\dagger}\left(\nabla_{\mu}-i\mathsf{g}\tau_{b}W^{b}_{\mu}-\frac{i}{2}\mathsf{g}'B_{\mu}\right)H_{0}+V_{eff}(H)\right\},
\end{equation}
where $V_{eff}(H_{0})=\mu_{eff}^{2}H_{0}^{\dagger}H_{0}-\lambda_{eff}\left(H_{0}^{\dagger}H_{0}\right)^{2}$, with
\begin{equation}
\mu_{eff}^{2}=\mu^{2}\frac{\displaystyle\int\limits_{-\infty}^{\infty}\mathrm{d}y e^{-6A}}{\displaystyle\int\limits_{-\infty}^{\infty}\mathrm{d}y e^{-4A}}\text{ and }\lambda_{eff}=\lambda\frac{\displaystyle\int\limits_{-\infty}^{\infty}\mathrm{d}y e^{-6A}}{\displaystyle\left(\int\limits_{-\infty}^{\infty}\mathrm{d}y e^{-4A}\right)^{2}}.
\end{equation}
After breaking the gauge symmetry, the Higgs field acquires a vacuum expectation value, driven by $\mu_{eff}$ and $\lambda_{eff}$, as
\begin{equation}
v=\frac{\mu_{eff}}{\sqrt{\lambda_{eff}}}=\frac{\mu}{\sqrt{\lambda}}\sqrt{\int\limits_{-\infty}^{\infty}\mathrm{d}y\left(f_{+}+f_{-}\right)^{4}},\label{vev}
\end{equation}	
and $\mathsf{m}_{k}v/\sqrt{2}$, from Eqs.~\eqref{massfermions} and~\eqref{vev}, gives the effective masses as measured by a four-dimensional observer.

The mechanism for the fine-tuning of the charged leptons masses can finally be explained. By assuming $F=\operatorname{sech}^{\mathsf{a}}\left(y\right)$, which is straightforwardly connected with
\begin{equation}
\mathsf{B}_{\theta}=-\frac{\mathsf{a}r}{\rho\mathsf{g}'}\tanh\left(y\right)\text{ and }\zeta=\frac{\mathsf{a}}{2}\ln\left[\operatorname{sech}\left(y\right)\right],\color{black}\label{gauge}
\end{equation}
and that $\mathsf{a}/{2}\leq2\rho/r<\mathsf{a}$, solely three normalizable fermionic zero modes can be identified, each of them associated with $k=-1$, $k=0$ and $k=1$, which are now to be labeled, respectively, as the electron, tauon and muon, i.e. $\mathsf{m}_{-1}=m_{e}$, $\mathsf{m}_{0}=m_{\tau}$ and $\mathsf{m}_{1}=m_{\mu}$. After some straightforward manipulations one finds
\begin{equation}
\frac{m_{\mu}}{m_{e}}=\frac{\displaystyle\int\limits^{\infty}_{-\infty}\mathsf{d}y\,\operatorname{sech}^{\mathsf{a}}\left(y\right)e^{2\frac{\rho}{r}y}\left(f_{+}+f_{-}\right)}{\displaystyle\int\limits^{\infty}_{-\infty}\mathsf{d}y\, \operatorname{sech}^{\mathsf{a}}\left(y\right)e^{-2\frac{\rho}{r}y}\left(f_{+}+f_{-}\right)},\label{mue}
\end{equation}
and
\begin{equation}
m_{\tau}=\left(m_{\mu}+m_{e}\right)\frac{\Gamma\left(\frac{\mathsf{a}-\frac{2\rho}{r}}{2}\right)\Gamma\left(\frac{\mathsf{a}+\frac{2\rho}{r}}{2}\right)\displaystyle\int\limits^{\infty}_{-\infty}\mathrm{d}y\,\operatorname{sech}^{\mathsf{a}}\left(y\right)f_{+}}{2\Gamma\left(\frac{\mathsf{a}}{2}\right)^{2}\displaystyle\int\limits^{\infty}_{-\infty}\mathsf{d}y\, \operatorname{sech}^{\mathsf{a}}\left(y\right)\cosh\left(2\frac{\rho}{r}y\right)f_{+}}.\color{black}\label{taumu}
\end{equation}
The largeness of the tauonic mass is an effect of canonical normalization and it is independent from the space-time asymmetry. If $\mathsf{a}$ is larger but of similar value to $2\rho/r$, then the wave functions of the electron and muon, which are not localized at the center, become spread out, while the wave function of the tauon gets localized at the center of the system of coordinates. After canonical normalization, the electronic and muonic masses pickup a very small term when compared with the tauonic term, thus explaining the largeness of the tauon mass. In this way, charged lepton mass constraints can be straightforwardly attained regardless of the asymmetry, since the tauon mass can be made as large as necessary, albeit not correctly valued. On the other hand, the relation between the electron and muon masses relies on the asymmetry of the warp factor. The wave functions of the electron and muon are a mirror of one each other, and the overlap between each and an asymmetric warp factor leads to different masses. Yet, not all asymmetric warp factors can conclude the correct masses, since a very light electron is only realized when $f_{+}-f_{-}$ goes to zero much faster than $f_{+}+f_{-}$ for positive $y$.


Being the SM particles represented by the zero modes, the massive modes are thus associated with beyond SM physics. From a phenomenological point of view, it is important to realize models that allow for the existence of a mass gap \cite{HERRERAAGUILAR2010}, between the zero and massive modes, in the spectrum of leptons. Then the energy scale at which the massive modes can be excited is fixed by this gap, and its existence is relevant for distinguishing the footprints of the massless modes, identified with stable four dimensional SM particles, from those coming from the massive modes, either discrete or continuous.  When no mass gap is present, then there exist several massive modes with masses small to the point of being indistinguishable from the massless ones. Specifically, for configurations with a non-trivial profile for the gauge and scalar field, like Eq.~\eqref{gauge}, the fermionic massive modes are solutions of Morse-Rosen equations\footnote{For further details see App. \ref{massivemodesleptons}.}, for which a discrete set of eigenstates can be identified, each associated with the mass eigenvalues
\begin{equation}
m^{+}_{jk}=\frac{2}{\rho}\frac{\sqrt{\left(j+1\right)\left(\mathsf{a}-j-1\right)}}{\left|\mathsf{a}-2j-2\right|}\sqrt{\left(\frac{\mathsf{a}}{2}-j-1\right)^{2}-\frac{k^{2}\rho^{2}}{r^{2}}},\label{discrete+}
\end{equation}
and
\begin{equation}
m^{-}_{jk}=\frac{1}{\rho}\frac{\sqrt{\left(j+1\right)\left(2\mathsf{a}-j-1\right)}}{\left|\mathsf{a}-j-1\right|}\sqrt{\left(\mathsf{a}-j-1\right)^{2}-\frac{k^{2}\rho^{2}}{r^{2}}},\label{discrete-}
\end{equation}
where $j$ is an integer, $0\leq j<\mathsf{a}/2-1$ and $m^{\pm}_{jk}$ references the discrete masses of positive and negative components of spinors, respectively. While continuous modes are related to masses 
\begin{equation}
m^{+}\geq \frac{1}{2\rho} \left|\mathsf{a}-\frac{2\rho\left|k\right|}{r}\right|\text{ and }m^{-}\geq \frac{1}{\rho} \left|\mathsf{a}-\frac{\rho\left|k\right|}{r}\right|,\color{black}
\end{equation}
therefore, a mass gap between the zero and massive modes is found whenever the zero modes are normalizable. 

To exemplify the mechanism in effect we propose a model for which $f_{+}=\operatorname{sech}^{l}\left(y\right)\cosh\left(oy\right)$ and $f_{-}=\operatorname{sech}^{l}\left(y\right)\sinh\left(oy\right)$, which, after substitution into Eq.~\eqref{bundlemetric}, leads to
\begin{equation}
\boldsymbol{g}=\operatorname{sech}^{2l}\left(y\right)e^{2oy}\left(\eta_{\mu\nu}\mathrm{d}x^{\mu}\mathrm{d}x^{\nu}+r^{2}\mathrm{d}\theta^{2}+\rho^{2}\mathrm{d}y^{2}\right).\color{black}\label{examplemetric}
\end{equation}

The fractions of the charged lepton masses associated with the metric \eqref{examplemetric}, calculated from Eqs.~\eqref{mue} and~\eqref{taumu}, become
\begin{equation}
\frac{m_{\mu}}{m_{e}}=\frac{\Gamma\left(\frac{l-o+\mathsf{a}-\frac{2\rho }{r}}{2}\right)\Gamma\left(\frac{l+o+\mathsf{a}+\frac{2\rho }{r}}{2}\right)}{\Gamma\left(\frac{l+o+\mathsf{a}-\frac{2\rho}{r}}{2}\right)\Gamma\left(\frac{l-o+\mathsf{a}+\frac{2\rho}{r}}{2}\right)},\label{mmue}\color{black}
\end{equation}
and
\begin{equation}
\frac{m_{\tau}}{m_{\mu}}=\frac{\Gamma\left(\frac{\mathsf{a}-\frac{2\rho}{r}}{2}\right)\Gamma\left(\frac{\mathsf{a}+\frac{2\rho}{r}}{2}\right) \Gamma\left(\frac{l+\mathsf{a}-o}{2}\right)\Gamma\left(\frac{l+\mathsf{a}+o}{2}\right)}{\Gamma\left(\frac{\mathsf{a}}{2}\right)^{2}\Gamma\left(\frac{l+\mathsf{a}-\frac{2\rho}{r}-o}{2}\right)\Gamma\left(\frac{l+\mathsf{a}+\frac{2\rho}{r}+o}{2}\right)}.\color{black}\label{mtaumu}
\end{equation}

\begin{figure}[!htb]
	\subfloat[]{\label{K2oide1}\includegraphics[scale=0.962]{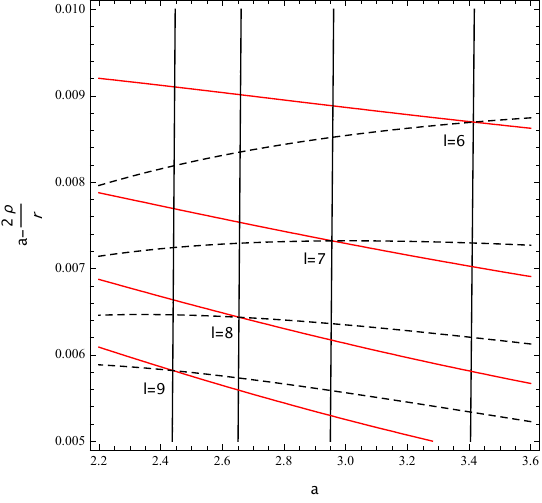}}$\;\;$
	\subfloat[]{\label{K2oide2}\includegraphics[scale=0.962]{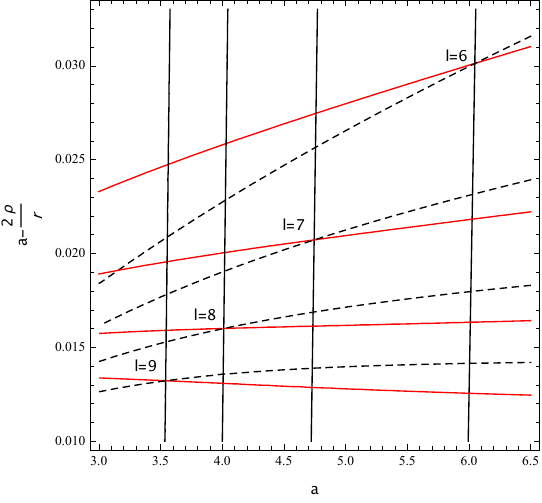}}
	\caption{(Color online) (a) The charged lepton mass spectrum associated with the metric \eqref{examplemetric} for $l-o=1$. (b) The charged lepton mass spectrum associated with the metric \eqref{examplemetric} for $l-o=2$. The solid black, solid red and dashed black lines represent the equations $m_{\mu}/m_{e}=206.768$, $m_{\tau}/m_{\mu}=16.817$ and $\mathcal{K}=2/3$, respectively. Results are for triple intersecting points at $l = 6,\,7,\,8,\,$ and $9$.} \label{K2oide}
\end{figure}

\begin{figure}[!htb]
	\subfloat[]{\label{lepton3D4}\includegraphics[scale=0.962]{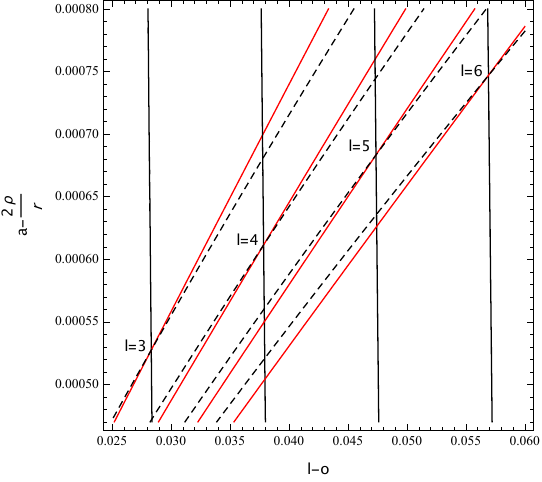}}$\;\;$
	\subfloat[]{\label{lepton3D3}\includegraphics[scale=0.962]{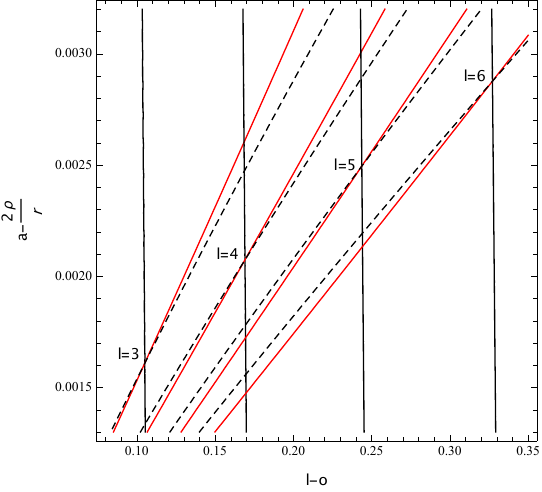}}
	\caption{(Color online) (a) The charged lepton mass spectrum associated with the metric \eqref{examplemetric} for $\mathsf{a}=1$. (b) The charged lepton mass spectrum associated with the metric \eqref{examplemetric} for $\mathsf{a}=2$. The solid black, solid red and dashed black lines represent the equations $m_{\mu}/m_{e}=206.768$, $m_{\tau}/m_{\mu}=16.817$ and $\mathcal{K}=2/3$, respectively. Results are for triple intersecting points at $l = 3,\,4,\,5,\,$ and $6$.}\label{lepton3D}
\end{figure}

If $o$ is an integer and $l-o$ is an even integer, then Eqs. \eqref{mmue} and \eqref{mtaumu} become polynomial equations. Particularly, for $o=2$ and $l=4$ one can solve Eqs. \eqref{mmue} and \eqref{mtaumu} analytically to find $\mathsf{a}=34.9562$ and $\mathsf{a}-2\rho/r=0.28488$. In fact, there are many combinations of $o$ and $l$ for which the lepton spectrum is achievable.
Fig. \ref{K2oide} depicts the values of $l$, $\mathsf{a}$ and $\mathsf{a}-2\rho/r$ with fixed $l-o$ for which the proper fractions of the masses and Koide's formula are realized for the metric Eq. \eqref{examplemetric}. Correspondently, Fig. \ref{lepton3D} depicts the values of $l$, $l-o$ and $\mathsf{a}-2\rho/r$ with fixed $\mathsf{a}$ for which the proper fractions of the masses and Koide's formula are realized for the metric Eq. \eqref{examplemetric}. The intersection of the curves in both Figs. \ref{K2oide} and \ref{lepton3D} are identified according to the parameter values that lead to the charged lepton mass spectrum. 

The lepton mass gap, between the continuous and massless modes, is $m\sim 10^{-2}/\rho$. A realistic mass gap, for small values of $l$, is thus achievable for $\rho\ll 10^{-18}\, \mathsf{m}$, since $m_{0}v/\sqrt{2}\approx m_{\tau}=1.7\, GeV$. On the other hand, for large values of $l$, tiny values of $1-o/l$ and $\mathsf{a}$ of the order of unity, a realistic mass gap is achievable for $\rho\ll 2^{l}\Gamma(\mathsf{a})/\left[l^{\mathsf{a}/2}\Gamma(\mathsf{a}/2)\right] 10^{-16}\mathsf{m}$, since $m_{0}v/\sqrt{2}\approx l^{\mathsf{a}/2}\Gamma(\mathsf{a}/2)/\left[2^{l}\Gamma(\mathsf{a})\right]GeV$. Therefore, for $l\sim 50$ the mass gap is achievable for $\rho\ll 1\, \mathsf{m}$. The mass gap for the Higgs field follows similar patterns.

To conclude, from a classical perspective, fermions are localized at the vicinity of an asymmetric conformally flat brane by the inclusion of a non-trivial bulk profile for the hypercharge gauge boson and a dark scalar field; and the proper mass fractions are a consequence of canonical normalization and the overlap between the fermionic wave functions with the asymmetric warp factor. The leptons of the SM are represented by the zero modes of six dimensional spinors, leading to a tower of linear independent modes driven by an integer value. After dimensional reduction, the setup reproduces the SM, with massless neutrinos, and there is no mismatch between the flavor and mass bases of the charged and neutral lepton sector. Otherwise, if a mechanism for neutrinos masses is included, the possibility of lepton flavor mixing and the non-conservation of lepton number apparently arises.
Noticing that flavor mixing is indeed relevant for the neutrino sector, fitting neutrino masses in intersecting braneworlds have been recently discussed \cite{Gemmill} in a framework where the near-tribimaximal mixing for neutrinos may arise naturally due to the structure of the Yukawa matrices.
Consistency with the quark and charged lepton mass matrices \cite{N1,N2,N3,N4} in combination with obtaining near-tribimaximal mixing fixes the Dirac neutrino mass matrix which is then driven by the seesaw mechanism for different choices of right-handed neutrino masses. In this formulation, running the obtained neutrino parameters down to the electroweak scale via the renormalization group equations leads to neutrino mass predictions \cite{Gemmill}.
Of course, given the intricate pattern of mass hierarchies and mixings between the different generations of quarks and neutrinos, our model was constrained to the charged lepton sector, where the effects of flavor mixing due to the coherent superpositions of flavor defined leptons, as assumed for neutrinos, are suppressed \cite{Nishi}.

A mechanism for explaining the spectrum of the charged lepton masses was therefore build upon two parameters: the gauge field strength, $\mathsf{a}$, and the ratio between the co-dimensional sizes, $\rho/r$, both related to a seminal $5+1$-dimensional braneworld scenario discussed in \cite{Gauy2022}.
In particular, from a model represented by metric Eq.~\eqref{examplemetric}, which was proposed for the sole reason of achieving analytical expressions, the proper fractions between the electron, muon and tauon masses were obtained, solely requiring that $\mathsf{a}-2\rho/r$ be tiny. Even when $\mathsf{a}$ is of the order of unity, the correct fractions for the masses can be realized when the warp factor presents a large asymmetry, which for Eq.~\eqref{examplemetric} is achievable when $1-o/l$ is tiny.
Furthermore, other asymmetric warp factors, in principle, also have the necessary structure to realize the needed mass spectrum, but an adjustment of the parameter values would be necessary. Even if surprising simple integrals have emerged when one assumed the metric as Eq.~\eqref{examplemetric}, other setups may not be so treatable as for finding analytical values for the masses. In this matter, relevant for next investigations, the same mechanism could also be employed for describing quark and neutrino mass hierarchy problems, since their diagonal masses should satisfy similar spectra. Of course, specific features related to quark or neutrino interactions, and their implications for the mass generation mechanism, turn the problem sufficiently more complex for such a preliminary analysis.

H.M.G. is grateful for the financial support provided by CNPq (Grant No. 141924/2019-5). The work of A.E.B. is supported by the Brazilian Agencies FAPESP (Grant No. 2023/00392-8) and CNPq (Grant No. 301485/2022-4).

	\appendix
	
	\section{Six-dimensional spinorial fields}\label{Spinorial}
	
	To investigate bulk spinorial matter, one first considers that fermion localization on brane-worlds
	is usually achieved when the 6-dimensional Dirac algebra is realized
	by the objects $\Gamma^{M}=\mathfrak{e}^{M}{}_{\bar{N}}\Gamma^{\bar{N}}$, where $\mathfrak{e}^{M}{}_{\bar{N}}$
	denotes a 6-dimensional vielbein,
	$\Gamma^{M}$ satisfy the Cliford relation $\left\{\Gamma^{M},\Gamma^{N}\right\}=2g^{MN}$, and $\Gamma^{\bar{N}}$
	are the gamma matrices in 6-dimensional flat space-time, for which the following representation\footnote{
		In representation \eqref{representation} the chirality matrix is diagonal, i.e.
		$
		\Gamma^{\bar{7}}=\begin{bmatrix}
		I^{4} &  0^{4}\\
		0^{4}  & -I^{4}
		\end{bmatrix}
		$.} is chosen
	\begin{equation}\label{representation}
	\Gamma^{\bar{\mu}}=\begin{bmatrix}
	0^{4} & \gamma^{\mu} \\
	\gamma^{\mu} & 0^{4} 
	\end{bmatrix},\text{ }\Gamma^{\bar{4}}=-\begin{bmatrix}
	0^{4} &  \gamma^{5}\\
	\gamma^{5} & 0^{4}
	\end{bmatrix},\text{ and }\Gamma^{\bar{5}}=i\begin{bmatrix}
	0^{4} & -I^{4} \\
	I^{4} & 0^{4}
	\end{bmatrix}.
	\end{equation}

	An $SU(2)_{L}\times U(1)$ action of a massless spinor, in $6$ dimensions, can be expressed as
	\begin{equation}
	S_{D}=\int\mathrm{d}^{6}x\sqrt{-g}\,\,\left[\overline{\mathsf{L}}^{\left(6\right)} {\Gamma}^{M}\left(\nabla_{M}-i\mathsf{g}\tau_{a}\mathsf{W}^{a}_{M}+\mathsf{g}'\frac{i}{2}\mathsf{B}_{M}\right)\mathsf{L}^{\left(6\right)}+\overline{\mathsf{e}}_{+}^{\left(6\right)}{\Gamma}^{M}\left(\nabla_{M}+i\mathsf{g}'\mathsf{B}_{M}+\zeta_{,M}\right)\mathsf{e}_{+}^{\left(6\right)}\right],\label{actionspinors}
	\end{equation}
	where $\nabla_{M}:=\partial_{M}+{\mathfrak{C}}_{M}$, $\zeta$ is some dark scalar field, $\mathsf{B}_{M}$ is the hypercharge gauge boson, $\mathsf{W}^{a}_{M}$ are the $SU(2)$ gauge bosons, $\mathsf{g}$ and $\mathsf{g}'$ are the $SU(2)$ and $U(1)$ couplings, $\tau_{a}=\sigma_{a}/2$ are the $SU(2)$ generators, and ${\mathfrak{C}}_{M}$ is the spin connection of $\left(\mathbb{E}^{6},\boldsymbol{g}\right)$, defined by
	\begin{equation}
	{\mathfrak{C}}_{M}=\frac{1}{4}\left\{-\frac{1}{2}{\mathfrak{e}_{M}}^{\bar{T}}\mathfrak{e}^{T\bar{R}}\mathfrak{e}^{Q\bar{S}}\partial_{\left[T\right.}\mathfrak{e}_{\left.Q\right]\bar{T}}+\frac{1}{2}\mathfrak{e}^{S\left[\bar{R}\right.}\partial_{\left[M\right.}{\mathfrak{e}_{\left.S\right]}}^{\left.\bar{S}\right]}\right\}\Gamma_{\bar{R}}\Gamma_{\bar{S}}.\label{spinconnection}
	\end{equation}
	The spin connection ${\mathfrak{C}}_{M}$, compatible with a conformally flat metric\footnote{That is a metric like $\boldsymbol{g}=e^{-2A}\left(\eta_{\mu\nu}\mathsf{d}x^{\mu}\mathsf{d}x^{\nu}+r^{2}\mathsf{d}\theta^{2}+\rho^{2}\mathsf{d}y^{2}\right)$.}, can be determined as follows
	\begin{equation}
	{\mathfrak{C}}_{\mu}={\mathfrak{A}}_{\mu}-\frac{1}{4}A_{,i}\hat{\mathfrak{e}}^{i\bar{j}}{\hat{\mathfrak{e}}_{\mu}}{}^{\bar{\nu}}\left(\Gamma_{\bar{\nu}}\Gamma_{\bar{j}}-\Gamma_{\bar{j}}\Gamma_{\bar{\nu}}\right),
	\end{equation}
	and
	\begin{equation}
	{\mathfrak{C}}_{i}=\frac{1}{4}\Gamma_{\bar{r}}\Gamma_{\bar{s}}\left\{-\frac{1}{2}{{\mathfrak{e}}_{i}}{}^{\bar{j}}{\mathfrak{e}}^{k\bar{r}}{\mathfrak{e}}^{l\bar{s}}\partial_{[k}{\mathfrak{e}}_{l]\bar{j}}+\frac{1}{2}{\mathfrak{e}}^{j[\bar{s}}\partial_{[j}{{\mathfrak{e}}_{i]}}{}^{\bar{r}]}
	\right\}={\mathfrak{B}}_{i}.
	\end{equation}
	Where ${\mathfrak{A}}_{\mu}$ and ${\mathfrak{B}}_{i}$ are the spin connections compatible with space-time $\left(\mathbb{M}^{4},\boldsymbol{\omega}\right)$ and internal space $\left(\mathbb{B}^{2},\boldsymbol{\sigma}\right)$, respectively. The left-handed leptons pair up to transform under $SU(2)$,
	\begin{equation}
	\mathsf{L}^{(6)}=\begin{Bmatrix}
	\nu^{(6)}_{+}\\\mathsf{e}^{(6)}_{+}
	\end{Bmatrix},
	\end{equation}
	where \begin{equation}
	\nu^{(6)}_{+}=\begin{bmatrix}
	\Psi^{L+}_{\nu}\\ \Psi^{R+}_{\nu}\\0\\0
	\end{bmatrix}\text{ and }\mathsf{e}^{(6)}_{+}=\begin{bmatrix}
	\Psi^{L+}_{\mathsf{e}}\\ \Psi^{R+}_{\mathsf{e}}\\0\\0
	\end{bmatrix}
	\end{equation} 
	represent the left-handed charged leptons and neutrinos, respectively. While right-handed leptons\footnote{Right-handed neutrinos have been excluded from the formalism since the interest in this work is in the charged lepton masses, and not on the mechanism of generating neutrinos masses.}, which are uncharged under $SU(2)$, are represented by
	\begin{equation}
	\mathsf{e}^{(6)}_{-}=\begin{bmatrix}
	0\\0\\\Psi^{L-}_{\mathsf{e}}\\ \Psi^{R-}_{\mathsf{e}}
	\end{bmatrix}.
	\end{equation}

	Varying the action $S_{d}$ with relation to $\overline{\mathsf{L}}^{(6)}$ or $\overline{\mathsf{e}}^{(6)}_{-}$ implies in the Dirac equation for $6$-dimensional curved space-time,
	\begin{equation}
	\left\{\hat{\Gamma}^{\mu}\mathcal{D}_{\mu}+e^{-A}\Gamma^{i}\left[\partial_{i}+{\mathfrak{B}}_{i}-2A_{,i}+\frac{1-\Gamma^{\bar{7}}}{2}\left(i\mathsf{g}'\mathsf{B}_{i}+\zeta_{,i}\right)-\frac{1+\Gamma^{\bar{7}}}{2}\left(i\mathsf{g}\tau_{a}\mathsf{W}^{a}_{i}-\frac{i}{2}\mathsf{g}'\mathsf{B}_{i}\right)\right]\right\}\Psi^{(6)}=0,\label{DiracD}
	\end{equation}
	where $\mathcal{D}_{\mu}=\partial_{\mu}+{\mathfrak{A}}_{\mu}+\frac{1-\Gamma^{\bar{7}}}{2}i\mathsf{g}'\mathsf{B}_{\mu}-\frac{1+\Gamma^{\bar{7}}}{2}\left(i\mathsf{g}\tau_{a}\mathsf{W}^{a}_{\mu}-\frac{i}{2}\mathsf{g}'\mathsf{B}_{\mu}\right)$ represents the usual covariant derivative of $\left(\mathbb{M}^{4},\boldsymbol{\omega}\right)$, $\Psi^{(6)}$ can represent either $\mathsf{L}^{(6)}$ or $\mathsf{e}^{(6)}_{-}$ and $\hat{\Gamma}^{\mu}=\hat{\mathfrak{e}}^{\mu}{}_{\bar{\nu}}\Gamma^{\bar{\nu}}=e^{-A}{\mathfrak{e}}^{\mu}{}_{\bar{\nu}}\Gamma^{\bar{\nu}}$, which is independent on the co-dimensions.
	
	Eq. \eqref{DiracD} is the Dirac equation for a general braneworld of any dimension\footnote{Which is true since ${\mathfrak{B}}_{i}$ can represent the spin connection of any co-dimensional space.}. For six dimensional spaces, a choice\footnote{There exists infinitely many choices of vielbeins compatible with some metric.} of a vielbein, compatible with a conformally flat metric, is simply $\mathfrak{e}_{4\bar{4}}=\mathfrak{e}_{4}{}^{\bar{4}}=\mathfrak{e}_{5\bar{5}}=\mathfrak{e}_{5}{}^{\bar{5}}=e^{-A}\text{ and }\mathfrak{e}^{4\bar{4}}=\mathfrak{e}^{4}{}_{\bar{4}}=\mathfrak{e}^{5\bar{5}}=\mathfrak{e}^{5}{}_{\bar{5}}=e^{A}.$
	Therefore, the Dirac equation for conformally flat six-dimensional braneworlds becomes
	\begin{equation}
	\left(\hat{\Gamma}^{\mu}\mathcal{D}_{\mu}+\mathcal{D}^{\bar{i}}\Gamma_{\bar{i}}\right)\Psi^{(6)}=0,\label{diracbrane6}
	\end{equation}
	where one defined the operators
	\begin{equation}
	\mathcal{D}^{\bar{5}}=\frac{1}{\rho}\left[\partial_{y}-\frac{5}{2}A_{,y}+\frac{1-\Gamma^{\bar{7}}}{2}\left(i\mathsf{g}'\mathsf{B}_{y}+\zeta_{,y}\right)-\frac{1+\Gamma^{\bar{7}}}{2}\left(i\mathsf{g}\tau_{a}\mathsf{W}^{a}_{y}-\frac{i}{2}\mathsf{g}'\mathsf{B}_{y}\right)\right],
	\end{equation}
	and
	\begin{equation}
	\mathcal{D}^{\bar{4}}=\frac{1}{r}\left[\partial_{\theta}-\frac{5}{2}A_{,\theta}+\frac{1-\Gamma^{\bar{7}}}{2}\left(i\mathsf{g}'\mathsf{B}_{\theta}+\zeta_{,\theta}\right)-\frac{1+\Gamma^{\bar{7}}}{2}\left(i\mathsf{g}\tau_{a}\mathsf{W}^{a}_{\theta}-\frac{i}{2}\mathsf{g}'\mathsf{B}_{\theta}\right)\right].
	\end{equation}
	Eq.~\eqref{diracbrane6} can be refined by separating the different chiralities\footnote{Any six dimensional spinor can be broken into $
		\Psi_{(6)}=\Psi^{+}_{(6)}+\Psi^{-}_{(6)}=\begin{bmatrix}
		\Psi_{+} \\
		0
		\end{bmatrix}+\begin{bmatrix}
		0 \\
		\Psi_{-}
		\end{bmatrix}=\begin{bmatrix}
		\Psi_{+} \\
		\Psi_{-}
		\end{bmatrix}
		$. Therefore, when $\Psi_{(6)}$ represents $\mathsf{L}^{(6)}$ or $\mathsf{e}^{(6)}_{-}$ it only contains $\Psi^{+}_{(6)}$ or $\Psi^{-}_{(6)}$, respectively.}, implying in two equations
	\begin{equation}
	i\hat{\gamma}^{\nu}\mathcal{D}^{-}_{\nu}\Psi_{-}-\mathcal{D}_{-}^{\bar{4}}\gamma^{5}\Psi_{-}-i\mathcal{D}_{-}^{\bar{5}}\Psi_{-}=0,\label{4dirac-}
	\end{equation}
	and
	\begin{equation}
	i\hat{\gamma}^{\nu}\mathcal{D}^{+}_{\nu}\Psi_{+}-\mathcal{D}_{-}^{\bar{4}}\gamma^{5}\Psi_{+}+i\mathcal{D}_{-}^{\bar{5}}\Psi_{+}=0,\label{4dirac+}
	\end{equation}
	where $\Psi_{\pm}$ are to be understood as four dimensional spinors, and
	\begin{equation}
	\mathcal{D}^{+}_{\mu}=\partial_{\mu}+{\mathfrak{A}}_{\mu}-\left(i\mathsf{g}\tau_{a}\mathsf{W}^{a}_{\mu}-\frac{i}{2}\mathsf{g}'\mathsf{B}_{\mu}\right),
	\end{equation}
	\begin{equation}
	\mathcal{D}^{-}_{\mu}=\partial_{\mu}+{\mathfrak{A}}_{\mu}+i\mathsf{g}'\mathsf{B}_{\mu},
	\end{equation}
	\begin{equation}
	\mathcal{D}_{+}^{\bar{5}}=\frac{1}{\rho}\left[\partial_{y}-\frac{5}{2}A_{,y}-i\mathsf{g}\tau_{a}\mathsf{W}^{a}_{y}+\frac{i}{2}\mathsf{g}'\mathsf{B}_{y}\right],
	\end{equation}
	\begin{equation}
	\mathcal{D}_{-}^{\bar{5}}=\frac{1}{\rho}\left[\partial_{y}+\zeta_{,y}-\frac{5}{2}A_{,y}+i\mathsf{g}'\mathsf{B}_{y}\right],
	\end{equation}
	\begin{equation}
	\mathcal{D}_{+}^{\bar{4}}=\frac{1}{r}\left[\partial_{\theta}-\frac{5}{2}A_{,\theta}-i\mathsf{g}\tau_{a}\mathsf{W}^{a}_{\theta}+\frac{i}{2}\mathsf{g}'\mathsf{B}_{\theta}\right],
	\end{equation}
	and
	\begin{equation}
	\mathcal{D}_{-}^{\bar{4}}=\frac{1}{r}\left[\partial_{\theta}+\zeta_{,\theta}-\frac{5}{2}A_{,\theta}+i\mathsf{g}'\mathsf{B}_{\theta}\right].
	\end{equation}

	\section{The quantum analogue problem for spinors}\label{wavefunctions}
	
	Suppose that the gauge and scalar fields satisfy
	\begin{equation}
	\mathsf{B}_{\mu}=\mathsf{B}_{\mu}\left(x^{\nu}\right)\text{, }\mathsf{W}^{a}_{\mu}=\mathsf{W}^{a}_{\mu}\left(x^{\nu}\right)\text{, }\mathsf{B}_{y}=\mathsf{W}^{a}_{i}=0\text{, }\mathsf{B}_{\theta}=\mathsf{B}_{\theta}\left(y\right)\text{ and }\zeta=\zeta\left(y\right).
	\end{equation}
	Thus a separation of variables technique,
	\begin{equation}
	\Psi_{\pm}=\sum_{m}\begin{bmatrix}
	L^{\pm}_{m}(\theta,v)\,\psi^{L}_{m\pm}(x^{\mu})\\R^{\pm}_{m}(\theta,v)\,\psi^{R}_{m\pm}(x^{\mu})
	\end{bmatrix}=\sum_{m}\left[L^{\pm}_{m}(\theta,v)\,\Psi^{L}_{m\pm}(x^{\mu})+R^{\pm}_{m}(\theta,v)\,\Psi^{R}_{m\pm}(x^{\mu})\right],
	\end{equation}
	can be employed for Eqs. \eqref{4dirac-} and \eqref{4dirac+}, implying in the equations
	\begin{equation}
	\hat{\gamma}^{\mu}\mathcal{D}^{\pm}_{\mu}
	\Psi^{L}_{m\pm}=m
	\Psi^{R}_{m\pm},\;\;\;\;\hat{\gamma}^{\mu}\mathcal{D}^{\pm}_{\mu}
	\Psi^{R}_{m\pm}=m
	\Psi^{L}_{m\pm},\label{ordinarydirac}
	\end{equation}
	and for the co-dimensional components
	\begin{equation}
	mR^{\pm}_{m}-\mathcal{D}_{\pm}^{\bar{4}}L^{\pm}_{m}\pm i\mathcal{D}_{\pm}^{\bar{5}}L^{\pm}_{m}=0,\label{massright}
	\end{equation}
	and
	\begin{equation}
	mL^{\pm}_{m}+\mathcal{D}_{\pm}^{\bar{4}}R^{\pm}_{m}\pm i\mathcal{D}_{\pm}^{\bar{5}}R^{\pm}_{m}=0.\label{massleft}
	\end{equation}
	A Schr\"odinger-like equation can be thus accomplished for the left and right-handed modes of spinors\footnote{For the zero modes one does not need any refinement, Eqs. \eqref{massright} and \eqref{massleft} can be solved as presented.}, respectively, as
	\begin{equation}
	m^{2}L^{+}_{m}+\mathcal{D}_{+}^{\bar{4}}\mathcal{D}_{+}^{\bar{4}}L^{+}_{m}-i\left[\mathcal{D}_{+}^{\bar{4}},\mathcal{D}_{+}^{\bar{5}}\right]L^{+}_{m}+\mathcal{D}_{+}^{\bar{5}}\mathcal{D}_{+}^{\bar{5}}L^{+}_{m}=0,\label{lefthanded+}
	\end{equation}
	\begin{equation}
	m^{2}R^{+}_{m}+\mathcal{D}_{+}^{\bar{4}}\mathcal{D}_{+}^{\bar{4}}R^{+}_{m}+i\left[\mathcal{D}_{+}^{\bar{4}},\mathcal{D}_{+}^{\bar{5}}\right]R^{+}_{m}+\mathcal{D}_{+}^{\bar{5}}\mathcal{D}_{+}^{\bar{5}}R^{+}_{m}=0,\label{righthanded+}
	\end{equation}
	\begin{equation}
	m^{2}L^{-}_{m}+\mathcal{D}_{-}^{\bar{4}}\mathcal{D}_{-}^{\bar{4}}L^{-}_{m}+i\left[\mathcal{D}_{-}^{\bar{4}},\mathcal{D}_{-}^{\bar{5}}\right]L^{-}_{m}+\mathcal{D}_{-}^{\bar{5}}\mathcal{D}_{-}^{\bar{5}}L^{-}_{m}=0,\label{lefthanded-}
	\end{equation}
	and
	\begin{equation}
	m^{2}R^{-}_{m}+\mathcal{D}_{-}^{\bar{4}}\mathcal{D}_{-}^{\bar{4}}R^{-}_{m}-i\left[\mathcal{D}_{-}^{\bar{4}},\mathcal{D}_{-}^{\bar{5}}\right]R^{-}_{m}+\mathcal{D}_{-}^{\bar{5}}\mathcal{D}_{-}^{\bar{5}}R^{-}_{m}=0,\label{righthanded-}
	\end{equation}
	where $\left[\mathcal{D}^{\bar{4}},\mathcal{D}^{\bar{5}}\right]=\mathcal{D}^{\bar{4}}\mathcal{D}^{\bar{5}}-\mathcal{D}^{\bar{5}}\mathcal{D}^{\bar{4}}$, which is generally non-null. On the other hand, the action for spinors is then given by
	\begin{align}
	S_{d}=&\sum_{\tilde{m}}\sum_{m}\int\mathrm{d}^{2}x\sqrt{\hat{\sigma}}e^{-5A}\overline{R}_{\tilde{m}}R_{m}\int\mathrm{d}^{4}x\sqrt{-\omega}\,
	\overline{\Psi}^{R}_{\tilde{m}+}
	\left(\hat{\gamma}^{\mu}\mathcal{D}_{\mu}
	\Psi^{R}_{m+}-m\,\Psi^{L}_{m+}\right)\nonumber
	\\
	&
	+\sum_{\tilde{m}}\sum_{m}\int\mathrm{d}^{2}x\sqrt{\hat{\sigma}}e^{-5A}\overline{L}_{\tilde{m}}L_{m}\int\mathrm{d}^{4}x\sqrt{-\omega}\,\overline{\Psi}^{L}_{\tilde{m}+}
	\left(\hat{\gamma}^{\mu}\mathcal{D}_{\mu}\Psi^{L}_{m+}-m\,\Psi^{R}_{m+}
	\right)\nonumber
	\\
	&+\sum_{\tilde{m}}\sum_{m}\int\mathrm{d}^{2}x\sqrt{\hat{\sigma}}e^{-5A}\overline{L}_{\tilde{m}}L_{m}\int\mathrm{d}^{4}x\sqrt{-\omega}\,
	\overline{\Psi}^{R}_{\tilde{m}-}
	\left(\hat{\gamma}^{\mu}\mathcal{D}_{\mu}
	\Psi^{R}_{m-}-m\,\Psi^{L}_{m-}\right)\nonumber
	\\
	&+\sum_{\tilde{m}}\sum_{m}\int\mathrm{d}^{2}x\sqrt{\hat{\sigma}}e^{-5A}\overline{R}_{\tilde{m}}R_{m}\int\mathrm{d}^{4}x\sqrt{-\omega}\,\overline{\Psi}^{L}_{\tilde{m}-}
	\left(\hat{\gamma}^{\mu}\mathcal{D}_{\mu}\Psi^{L}_{m-}-m\,\Psi^{R}_{m-}
	\right)
	\end{align}
	which corresponds to the same problem driven by Eq. \eqref{ordinarydirac}, if and only if the modes are
	\begin{enumerate}
		\item Normalizable, i.e. $\displaystyle\int\mathrm{d}^{2}x\sqrt{\hat{\sigma}}e^{-5A}\overline{R}_{m}R_{m}=1\text{ and }\displaystyle\int\mathrm{d}^{2}x\sqrt{\hat{\sigma}}e^{-5A}\overline{L}_{m}L_{m}=1,\;\;\forall m$;
		\item Orthogonal, i.e. $\displaystyle\int\mathrm{d}^{2}x\sqrt{\hat{\sigma}}e^{-5A}\overline{R}_{\tilde{m}}R_{m}=0$ and $\displaystyle\int\mathrm{d}^{2}x\sqrt{\hat{\sigma}}e^{-5A}\overline{L}_{\tilde{m}}L_{m}=0\text{ if }\tilde{m}\neq m$.
	\end{enumerate}
	When the above conditions are satisfied then the action reads
	\begin{equation}
	S_{d}=\sum_{m}\int\mathrm{d}^{4}x\sqrt{-\omega}\,
	\overline{\Psi}^{+}_{m}
	\left(\hat{\gamma}^{\mu}\mathcal{D}_{\mu}
	\Psi^{+}_{m}-m\,\Psi^{+}_{m}\right)+\sum_{m}\int\mathrm{d}^{4}x\sqrt{-\omega}\,
	\overline{\Psi}^{-}_{m}
	\left(\hat{\gamma}^{\mu}\mathcal{D}_{\mu}
	\Psi^{-}_{m}-m\,\Psi^{-}_{m}\right).
	\end{equation}
	
	Eqs. \eqref{massright} and \eqref{massleft} can be readily integrated for the zero modes ($m=0$), leading to
	\begin{align}
	L^{+}_{0}&=e^{\frac{5A}{2}}\sum_{k}L^{+}_{0k}=\sum_{k}{C}^{L+}_{k}e^{ik\theta}e^{\frac{5A}{2}}e^{\frac{k\rho y}{r}}e^{-\frac{e\rho}{2r}\int\mathsf{B}_{\theta}\mathrm{d}y},\label{L+}
	\\
	R^{+}_{0}&=e^{\frac{5A}{2}}\sum_{k}R^{+}_{0k}=\sum_{k}{C}^{R+}_{k}e^{-ik\theta}e^{\frac{5A}{2}}e^{\frac{k\rho y}{r}}e^{\frac{e\rho}{2r}\int\mathsf{B}_{\theta}\mathrm{d}y},\label{R+}
	\\
	L^{-}_{0}&=e^{\frac{5A}{2}}\sum_{k}L^{-}_{0k}=\sum_{k}{C}^{L-}_{k}e^{-ik\theta}e^{\frac{5A}{2}}e^{\frac{k\rho y}{r}}e^{\frac{e\rho}{r}\int\mathsf{B}_{\theta}\mathrm{d}y}e^{-\zeta},\label{L-}
	\end{align}
	and
	\begin{equation}
	\\
	R^{-}_{0}=e^{\frac{5A}{2}}\sum_{k}R^{-}_{0k}=\sum_{k}{C}^{R-}_{k}e^{ik\theta}e^{\frac{5A}{2}}e^{\frac{k\rho y}{r}}e^{-\frac{e\rho}{r}\int\mathsf{B}_{\theta}\mathrm{d}y}e^{-\zeta},\label{R-}
	\end{equation}
	where $k\in\mathbb{N}$ and ${C}^{R(L)\pm}_{k}$ are constants. And the zero mode normalization reads
	\begin{align}
	2\pi\left({C}^{L+}_{k}\right)^{2}\int\mathrm{d}y\,e^{\frac{2k\rho y}{r}}e^{-\frac{e\rho}{r}\int\mathsf{B}_{\theta}\mathrm{d}y}&=1,
	\\
	2\pi\left({C}^{R+}_{k}\right)^{2}\int\mathrm{d}y\,e^{\frac{2k\rho y}{r}}e^{\frac{e\rho}{r}\int\mathsf{B}_{\theta}\mathrm{d}y}&=1,
	\\
	2\pi\left({C}^{L-}_{k}\right)^{2}\int\mathrm{d}y\,e^{\frac{2k\rho y}{r}}e^{-2\zeta}e^{\frac{2e\rho}{r}\int\mathsf{B}_{\theta}\mathrm{d}y}&=1,
	\end{align}
	and
	\begin{equation}
	2\pi\left({C}^{R-}_{k}\right)^{2}\int\mathrm{d}y\,e^{\frac{2k\rho y}{r}}e^{-2\zeta}e^{-\frac{2e\rho}{r}\int\mathsf{B}_{\theta}\mathrm{d}y}=1.
	\end{equation}
	Clearly $L^{+}_{0}$ and $R^{-}_{0}$ can not be normalized at the same time as $R^{+}_{0}$ and $L^{-}_{0}$. Normalizable zero modes are thus written as
	\begin{equation}
	\Psi^{(6)}_{0}=e^{\frac{5A}{2}}\begin{bmatrix}
	\sum_{k}L^{+}_{0k}(\theta,y)\,\psi^{L+}_{0k}(x^{\mu})\\0
	\\0\\\sum_{k}R^{-}_{0k}(\theta,y)\,\psi^{R-}_{0k}(x^{\mu})
	\end{bmatrix},
	\end{equation}
	where $L^{+}_{0k}$ and $R^{-}_{0k}$ are the zero mode wave functions described by Eqs. \eqref{L+} and \eqref{R-}.
	\section{On the inclusion of a mass perturbation}
	Fermions can be described by the action
	\begin{equation}
	\overline{S}_{d}=\int\mathrm{d}^{6}x\sqrt{-g}\,\left(\overline{\mathsf{L}}^{(6)}\Gamma^{M}\mathsf{D}_{M}\mathsf{L}^{(6)}+\overline{\mathsf{e}}_{-}^{(6)}\Gamma^{M}\mathsf{D}_{M}\mathsf{e}_{-}^{(6)}\right)-m_{0}\int\mathrm{d}x^{6}\sqrt{-g}\,\left(\overline{\mathsf{L}}^{(6)}H{\mathsf{e}}_{-}^{(6)}+\overline{\mathsf{e}}_{-}^{(6)}H^{\dagger}{\mathsf{L}}^{(6)}\right),\label{sixdirac}
	\end{equation}
	where the mass term coupling $m_{0}$ should be small if compared to the scale of the brane, $r$ and $\rho$, and should be treated as perturbation, presenting negligible effects in the wave functions. This can be justified directly from the action
	\begin{align}
	\overline{S}_{d}=&\int\mathrm{d}^{6}x\sqrt{-g}\,\left[\overline{\mathsf{L}}^{(6)}\left(\Gamma^{\mu}\mathsf{D}_{\mu}+\frac{1}{\rho}\Gamma^{y}\mathsf{D}_{y}+\frac{1}{r}\Gamma^{\theta}\mathsf{D}_{\theta}\right)\mathsf{L}^{(6)}+\overline{\mathsf{e}}_{-}^{(6)}\left(\Gamma^{\mu}\mathsf{D}_{\mu}+\frac{1}{\rho}\Gamma^{y}\mathsf{D}_{y}+\frac{1}{r}\Gamma^{\theta}\mathsf{D}_{\theta}\right)\mathsf{e}_{-}^{(6)}\right]\nonumber
	\\
	&\qquad\qquad\qquad\qquad\qquad\qquad\qquad\qquad\qquad\qquad\qquad\qquad\qquad-m_{0}\int\mathrm{d}x^{6}\sqrt{-g}\,\left(\overline{\mathsf{L}}^{(6)}H{\mathsf{e}}_{-}^{(6)}+\overline{\mathsf{e}}_{-}^{(6)}H^{\dagger}{\mathsf{L}}^{(6)}\right),
	\end{align}
	which implies that the co-dimensional portion of the action is of the order of $1/r$ or $1/\rho$, while the rest is of the order of $m_{0}$. Thus the fermionic co-dimensional wave functions even with the perturbative term can be determined exactly as the previous sections, and should be described by the Eqs. \eqref{lefthanded+}, \eqref{righthanded+}, \eqref{lefthanded-} and \eqref{righthanded-}.
	
	To realize the effective theory that is observed in four dimensions one substitutes the zero modes, Eqs. \eqref{L+} to \eqref{R-}, into action \eqref{sixdirac}. The mass term is thus described by
	\begin{align}
	\int\mathrm{d}x^{6}\sqrt{-g}\,\left(\overline{\mathsf{L}}^{(6)}H{\mathsf{e}}_{-}^{(6)}+\overline{\mathsf{e}}_{-}^{(6)}H^{\dagger}{\mathsf{L}}^{(6)}\right)&=\int\mathrm{d}x^{6}\sqrt{-g}\,\left(\mathsf{L}^{(6)}{}^{\dagger}\Gamma^{0}H{\mathsf{e}}_{-}^{(6)}+\mathsf{e}_{-}^{(6)}{}^{\dagger}\Gamma^{0}H^{\dagger}{\mathsf{L}}^{(6)}\right)\nonumber
	\\
	&= \sum\limits_{p,q,m,n}\int\mathrm{d}x^{6}\sqrt{-g}\begin{bmatrix}
	\overline{{L}^{+}_{0p}}\psi^{L+\dagger}_{0p}&
	0&
	0&
	0
	\end{bmatrix}
	\begin{bmatrix}
	0^{4} & \gamma^{0} \\
	\gamma^{0} & 0^{4} 
	\end{bmatrix}
	\Phi H_{0}\begin{bmatrix}
	0\\
	0\\
	0\\
	{R}^{-}_{0n}\psi^{R-}_{0n}
	\end{bmatrix}\nonumber
	\\
	&+ \sum\limits_{p,q,m,n}\int\mathrm{d}x^{6}\sqrt{-g}\begin{bmatrix}
	0&
	0&
	0&
	\overline{{R}^{-}_{0q}}\psi^{R-\dagger}_{0q}
	\end{bmatrix}
	\begin{bmatrix}
	0^{4} & \gamma^{0} \\
	\gamma^{0} & 0^{4} 
	\end{bmatrix}
	\Phi H_{0}\begin{bmatrix}
	{L}^{+}_{0m}\psi^{L+}_{0m}\\
	0\\
	0\\
	0
	\end{bmatrix}\nonumber
	\\
	&=\sum_{p}\int\mathrm{d}y^{2}\sqrt{\hat{\sigma}}e^{-6A}\Phi\overline{{L}^{+}_{0p}}{R}^{-}_{0p}\int\mathrm{d}x^{4}\overline{\Psi}^{L+}_{0p}H_{0}{\Psi}^{R-}_{0p}
	\\
	&+\sum_{k}\int\mathrm{d}y^{2}\sqrt{\hat{\sigma}}e^{-6A}\Phi\overline{{R}^{-}_{0k}}{L}^{+}_{0k}\int\mathrm{d}x^{4}\overline\Psi^{R-}_{0k}H_{0}^{\dagger}\Psi^{L+}_{0k},
	\end{align}
	where $\Psi^{L+}_{0k}$ can represent both neutrinos and charged leptons, one has already employed the fact that ${{L}^{\pm}_{0k}}$ is orthogonal to ${R}^{\pm}_{0p}$ if $k\neq p$, assumed that $\Phi$ is real valued and excluded the non-normalizable states, i.e. ${L}^{-}$ and ${R}^{+}$. Finally, resuming from the total action and canonically normalizing it, one finds
	\begin{align}
	S_{D}^{\left(4\right)}=&\sum_{k}\int\mathrm{d}^{4}x\Bigg\{i\,\overline{\mathsf{L}}_{k}\,\gamma^{\mu}\Bigg[\nabla_{\mu}-\left(i\mathsf{g}\tau_{a}\mathsf{W}^{a}_{\mu}-\frac{i}{2}\mathsf{g}'\mathsf{B}_{\mu}\right)\Bigg]\,\mathsf{L}_{k}+i\,\overline{\mathsf{e}}_{Rk}\gamma^{\mu}\Bigg(\nabla_{\mu}+i\mathsf{g}'\mathsf{B}_{\mu}\Bigg)\mathsf{e}_{Rk}\nonumber
	\\
	&\qquad\qquad\qquad\qquad\qquad\qquad\qquad\qquad\qquad\qquad\qquad\qquad\qquad\qquad\quad-m_{k}\left[\,\overline{\mathsf{L}}_{k}\,H_{0}\,\mathsf{e}_{Rk}+\,\overline{\mathsf{e}}_{Rk}\,H^{\dagger}_{0}\,\mathsf{L}_{k}\right]\Bigg\},\label{dirac4}
	\end{align}
	where
	\begin{equation}
	\mathsf{L}_{k}=\begin{Bmatrix}
	\nu_{Lk}\\\mathsf{e}_{Lk}
	\end{Bmatrix}\text{, }
	\nu_{Lk}=\begin{bmatrix}
	\psi^{L+}_{\nu 0k}\\0
	\end{bmatrix}\text{, }\mathsf{e}_{Lk}=\begin{bmatrix}
	\psi^{L+}_{\mathsf{e} 0k}\\0
	\end{bmatrix}\text{, }\mathsf{e}_{Rk}=\begin{bmatrix}
	0\\ \psi^{R-}_{\mathsf{e}0k}
	\end{bmatrix}
	\end{equation}
	and
	\begin{equation}
	m_{k}=\frac{\displaystyle m_{0}\int\mathrm{d}y^{2}\sqrt{\hat{\sigma}}e^{-6A}\Phi\overline{{L}^{+}_{0k}}{R}^{+}_{0k}}{\displaystyle\sqrt{\int\mathrm{d}y^{2}\sqrt{\hat{\sigma}}e^{-5A}\overline{L^{+}_{0k}}{L}^{+}_{0k}}\sqrt{\int\mathrm{d}y^{2}\sqrt{\hat{\sigma}}e^{-5A}\overline{R^{+}_{0k}}{R}^{+}_{0k}}\sqrt{\int\mathrm{d}y^{2}\sqrt{\hat{\sigma}}e^{-4A}\Phi^{2}}}.\label{massdirac}
	\end{equation}
	Action \eqref{dirac4} described $k$-many usual four dimensional Dirac spinors each with a mass described by Eq. \eqref{massdirac}.
	
	\section{Massive Modes: The Leptons}\label{massivemodesleptons}
	The last important aspect to realizing realistic models is achieving a mass gap between the zero mode and the massive modes. To this end one assumes the gauge field
	\begin{equation}
	\mathsf{B}_{\theta}=-\mathsf{q}\tanh\left(y\right),
	\end{equation}
	which once applied to Eqs. \eqref{lefthanded+}, \eqref{righthanded+}, \eqref{lefthanded-} and \eqref{righthanded-}, after a rescaling of the wave functions $L^{+}_{m}=e^{5A/2}\sum_{k} e^{ik\theta}\mathsf{L}^{+}_{mk}$ and $L^{-}_{m}=e^{5A/2}e^{-\zeta}\sum_{k} e^{ik\theta}\mathsf{L}^{-}_{mk}$ (the same is true for $R^{+}_{m}$ and $R^{-}_{m}$), implies in four Morse-Rosen equations,
	\begin{equation}
	-\mathsf{L}^{+}_{mk,yy}+\left[-\frac{\lambda}{2}\left(\frac{\lambda}{2}+1\right)\operatorname{sech}^{2}\left(y\right)-ks\tanh\left(y\right)\right]\mathsf{L}^{+}_{mk}=\left(m^{2}\rho^{2}-\frac{k^{2}s^{2}}{\lambda^{2}}-\frac{\lambda^{2}}{4}\right)\mathsf{L}^{+}_{mk},\label{lefttrivialschroedinger+}
	\end{equation}
	\begin{equation}
	-\mathsf{R}^{+}_{mk,yy}+\left[-\frac{\lambda}{2}\left(\frac{\lambda}{2}-1\right)\operatorname{sech}^{2}\left(y\right)-ks\tanh\left(y\right)\right]\mathsf{R}^{+}_{mk}=\left(m^{2}\rho^{2}-\frac{k^{2}s^{2}}{\lambda^{2}}-\frac{\lambda^{2}}{4}\right)\mathsf{R}^{+}_{mk},\label{righttrivialschroedinger+}
	\end{equation}
	\begin{equation}
	-\mathsf{L}^{-}_{mk,yy}+\left[-\lambda\left(\lambda+1\right)\operatorname{sech}^{2}\left(y\right)-2ks\tanh\left(y\right)\right]\mathsf{L}^{-}_{mk}=\left(m^{2}\rho^{2}-\frac{k^{2}s^{2}}{\lambda^{2}}-\lambda^{2}\right)\mathsf{L}^{-}_{mk},\label{lefttrivialschroedinger-}
	\end{equation}
	and
	\begin{equation}
	-\mathsf{R}^{-}_{mk,yy}+\left[-\lambda\left(\lambda-1\right)\operatorname{sech}^{2}\left(y\right)-2ks\tanh\left(y\right)\right]\mathsf{R}^{-}_{mk}=\left(m^{2}\rho^{2}-\frac{k^{2}s^{2}}{\lambda^{2}}-\lambda^{2}\right)\mathsf{R}^{-}_{mk}.\label{righttrivialschroedinger-}
	\end{equation}
	where $\lambda=\frac{\mathsf{g}'\rho \mathsf{q}}{r}$, $s=\frac{\rho\lambda}{r}$ and $\mathsf{q}$ is a real valued constant. Eqs. \eqref{lefttrivialschroedinger-} and \eqref{righttrivialschroedinger-} are equivalent to Eqs. \eqref{lefttrivialschroedinger+} and \eqref{righttrivialschroedinger+}, one just needs to switch $\lambda$ for $\lambda/2$. Therefore, from now on, only Eqs. \eqref{lefttrivialschroedinger-} and \eqref{righttrivialschroedinger-} shall be considered for further investigation. The quantum mechanical potential associated with left and right-handed spinors are, respectively,
	\begin{equation}
	U_{R}=\left[-\lambda\left(\lambda-1\right)\operatorname{sech}^{2}\left(y\right)-2ks \tanh\left(y\right)\right],\label{trivialpotentialright}
	\end{equation}
	and
	\begin{equation}
	U_{L}=\left[-\lambda\left(\lambda+1\right)\operatorname{sech}^{2}\left(y\right)-2ks \tanh\left(y\right)\right].\label{trivialpotentialleft}
	\end{equation}
	
\begin{figure}[!htb]
		\subfloat[]{\includegraphics[scale=0.95]{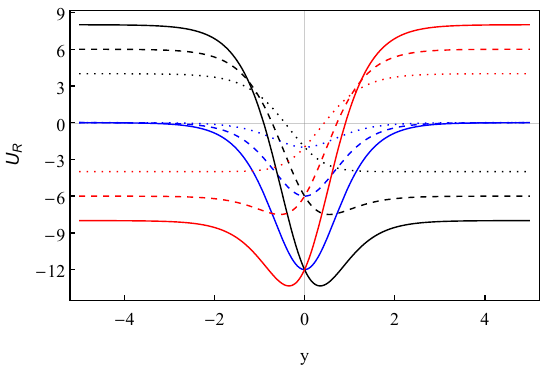}}$\;\;$
		\subfloat[]{\includegraphics[scale=0.95]{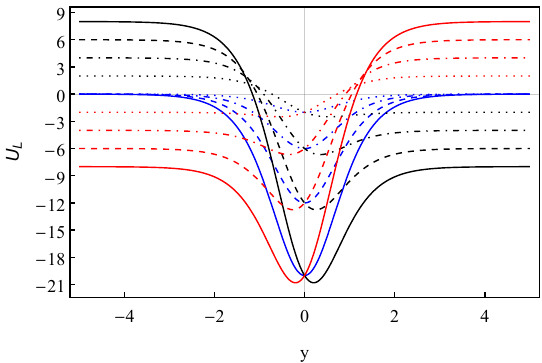}}
		\caption{(Color online) (a) The quantum mechanical potential associated with right-handed spinors, $U_{R}$, for $\lambda=4$ (solid lines), $\lambda=3$ (dashed lines) and $\lambda=2$ (dotted lines). (b)  The quantum mechanical potential associated with left-handed spinors, $U_{L}$, for $\lambda=4$ (solid lines), $\lambda=3$ (dashed lines), $\lambda=2$ (dot-dashed lines) and $\lambda=1$ (dotted lines). The plots are for $k=0$ (blue lines), $k=1$ (black lines) and $k=-1$ (red lines), with $s=1$.}\label{Trivialquantumpotential}
	\end{figure}

	The potentials $U_{R}$ and $U_{L}$, Eqs. \eqref{trivialpotentialright} and \eqref{trivialpotentialleft}, have an associated global minima\footnote{It is assumed that $\lambda>0$.} if $\lambda>1$ and $\lambda\left(\lambda-1\right)>s\left|k\right|$, and $\lambda>0$ and $\lambda\left(\lambda+1\right)>s\left|k\right|$, respectively. Thus creating the conditions for producing bound states for all real valued $s$ and integer $k$.
	The quantum mechanical potential, Eqs. \eqref{trivialpotentialright} and \eqref{trivialpotentialleft}, of the left and right-handed equations are depicted in Fig. \ref{Trivialquantumpotential}. 

	The general solution of Eqs. \eqref{lefttrivialschroedinger-} and \eqref{righttrivialschroedinger-} are, respectively,
	\begin{align}
	\mathsf{R}^{-}_{mk}=& c_1\operatorname{sech}^{q}\left(y\right)e^{py} {}_2F_1\left(q+1-\lambda,\lambda +q;q-p+1;\frac{e^{-y}\operatorname{sech} (y)}{2} \right)\nonumber
	\\
	&\qquad\qquad\qquad\qquad\qquad\qquad\qquad\qquad\;+c_2\operatorname{sech}^{p}\left(y\right)e^{qy} \, _2F_1\left(-\lambda +p+1,\lambda+p;p-q+1;\frac{e^{-y}\operatorname{sech} (y)}{2} \right),\label{solutiontrivialright}
	\end{align}
	and
	\begin{align}
	\mathsf{L}^{-}_{mk}=& d_1\operatorname{sech}^{q}\left(y\right)e^{py} {}_2F_1\left(q-\lambda,\lambda+1 +q;q-p+1;\frac{e^{-y}\operatorname{sech} (y)}{2} \right)\nonumber
	\\
	&\qquad\qquad\qquad\qquad\qquad\qquad\qquad\qquad\;+d_2\operatorname{sech}^{p}\left(y\right)e^{qy} \, _2F_1\left(-\lambda +p,\lambda+1+p;p-q+1;\frac{e^{-y}\operatorname{sech} (y)}{2} \right),\label{solutiontrivialleft}
	\end{align}
	where $q=\frac{\sqrt{\sqrt{\epsilon ^2-4k^2 s^2}-\epsilon }}{\sqrt{2}}$ and $p=\operatorname{sign}\left(k\right)\frac{\sqrt{-\epsilon -\sqrt{\epsilon ^2-4k^2 s^2}}}{\sqrt{2}}$. If $\epsilon\geq-2\left|k\right|s$ or $m\geq \frac{1}{\rho}\left|\lambda-\frac{\rho\left|k\right|}{r}\right|$ then Eqs. \eqref{solutiontrivialright} and \eqref{solutiontrivialleft} correspond to the propagating modes. Otherwise, for $\epsilon<-\left|k\right|s$, the solutions \eqref{solutiontrivialright} and \eqref{solutiontrivialleft} lead to the bound states
	\begin{equation}
	\mathsf{R}^{-}_{mk}= c_1\operatorname{sech}^{\lambda-j-1}\left(y\right)e^{\frac{ks}{\left(\lambda-j-1\right)}y} {}_2F_1\left(-j,2\lambda-j-1;\lambda+\frac{ks}{\left(\lambda-j-1\right)}-j;\frac{e^{y}\operatorname{sech} (y)}{2} \right),\label{boundtrivialright-}
	\end{equation}
	and
	\begin{equation}
	\mathsf{L}^{-}_{mk}= d_1\operatorname{sech}^{\lambda-j-1}\left(y\right)e^{\frac{ks}{\left(\lambda-j-1\right)}y} {}_2F_1\left(-j-1,2\lambda-j;\lambda+\frac{ks}{\left(\lambda-j-1\right)}-j;\frac{e^{y}\operatorname{sech} (y)}{2} \right),\label{boundtrivialleft-}
	\end{equation}
	both associated with the mass eigenvalues
	\begin{equation}
	m^{-}_{jk}=\frac{1}{\rho}\frac{\sqrt{\left(j+1\right)\left(2\lambda-j-1\right)}}{\lambda\left(\lambda-j-1\right)}\sqrt{\lambda^{2}\left(\lambda-j-1\right)^{2}-k^{2}s^{2}},\label{massdiscrete-}
	\end{equation}
	where\footnote{With the exception of the zero mode that is constructed from Eq. \eqref{boundtrivialleft-} with $j=-1$.} $j$ is a natural number. In this way there is always a mass gap between the zero and massive modes, be it the discrete or continuous modes. Clearly the solutions and spectrum of mass of $\mathsf{L}^{+}_{mk}$ and $\mathsf{R}^{+}_{mk}$ are similar to Eqs. \eqref{boundtrivialright-}, \eqref{boundtrivialleft-} and \eqref{massdiscrete-}, but with $\lambda\rightarrow\lambda/2$.
	
	\section{Massive Modes: The Higgs Field}
	In the same line of reasoning, the Higgs field must also have a mass gap between the zero and massive modes. But the massive modes of the Higgs depends upon the choice of metric. Particularly, for the metric considered in this paper,
	\begin{equation}
	\boldsymbol{g}=\operatorname{sech}^{2l}\left(y\right)e^{2oy}\left(\eta_{\mu\nu}\mathrm{d}x^{\mu}\mathrm{d}x^{\nu}+r^{2}\mathrm{d}\theta^{2}+\rho^{2}\mathrm{d}y^{2}\right).\color{black}
	\end{equation}
	and the gauge boson is once again
	\begin{equation}
	\mathsf{B}_{\theta}=-\mathsf{q}\tanh\left(y\right),
	\end{equation}
	the bulk profile of the Higgs field follows from the equation
	\begin{equation}
	-\phi_{,yy}
	+\left[-2l\left(2l+1\right)\operatorname{sech}^{2}\left(y\right)-\left(8ol-ks\right)\tanh(y)+4l^{2}+4o^{2}+\frac{k^{2}s^{2}}{\lambda^{2}}\right]\phi=m^{2}\rho^{2}\phi\label{higgsbulk}
	\end{equation}
	where $\lambda=\frac{\mathsf{g}'\rho \mathsf{q}}{r}$ and $s=\frac{\lambda\rho}{r}$. The general solution of Eq. \eqref{higgsbulk} takes the form
	\begin{align}
	\phi_{kj}=& c_1\operatorname{sech}^{q}\left(y\right)e^{py} {}_2F_1\left(q-2l,2l+1 +q;q-p+1;\frac{e^{-y}\operatorname{sech} (y)}{2} \right)\nonumber
	\\
	&\qquad\qquad\qquad\qquad\qquad\qquad\qquad\qquad\;+c_2\operatorname{sech}^{p}\left(y\right)e^{qy} \, _2F_1\left(p-2l ,2l+1+p;p-q+1;\frac{e^{-y}\operatorname{sech} (y)}{2} \right),\label{higgswave}
	\end{align}
	where $q=\frac{\sqrt{\sqrt{\epsilon ^2-\left(8ol-ks\right)^2}-\epsilon }}{\sqrt{2}}$, $p=\frac{\left|8ol-ks\right|}{2q}$ and $\epsilon=m^{2}\rho^{2}-4l^{2}-4o^{2}-\frac{k^{2}s^{2}}{\lambda^{2}}$. If $\epsilon\geq-\left|8ol-ks\right|$ then Eq. \eqref{higgswave} corresponds to the propagating modes. Otherwise, for $\epsilon<-\left|8ol+ks\right|$, the solution \eqref{higgswave} leads to the bound states
	\begin{equation}
	\phi_{kj}(y)=c e^{\frac{\left|8ol-ks\right|}{2\left(2l-j\right)}y}\operatorname{sech}^{2l-j}\left(y\right) \, _2F_1\left(-j,4l+1-j;2l+1-j-\frac{\left|4ol-ks\right|}{2l-j};\frac{e^{-y}\operatorname{sech}\left(y\right)}{2}\right).
	\end{equation}
	both associated with the mass eigenvalues
	\begin{equation}
	m^{2}\rho^{2}=4l^{2}+4o^{2}+\frac{k^{2}s^{2}}{\lambda^{2}}-\left(2l-j\right)^{2}-\frac{\left(8ol-ks\right)^{2}}{4\left(2l-j\right)^{2}},\label{higgsdiscrete}
	\end{equation}
	where $j$ is a natural number. In this way there is always a mass gap between the zero and massive modes, be it a discrete or continuous mode.

	\footnotesize
	

\end{document}